\documentclass[floatfix,aps,pra,superscriptaddress,twocolumn,longbibliography]{revtex4-2}

\usepackage[english]{babel}

\usepackage{graphicx}
\usepackage[colorlinks=true, allcolors=blue]{hyperref}
\usepackage{physics}
\usepackage{amsmath,amsfonts,amssymb,amsthm}
\usepackage{enumitem}
\usepackage[utf8]{inputenc}
\usepackage{lipsum}
\usepackage{ulem}

\usepackage[dvipsnames]{xcolor}

\usepackage{natbib}
\usepackage{url}
\hypersetup{colorlinks=true, urlcolor=blue, linkcolor=blue, citecolor=blue}

\begin{document}

\title{Dynamical invariant based shortcut to equilibration in open quantum systems}

\author{Mohamed Boubakour}
\email{mohamed.boubakour@oist.jp}
\affiliation{Quantum Systems Unit, Okinawa Institute of Science and Technology Graduate University, Okinawa 904-0495, Japan}
\author{Shimpei Endo}
\email{shimpei.endo@uec.ac.jp}
\affiliation{Department of Engineering Science, The University of Electro-Communications, Tokyo 182-8585, Japan}
\affiliation{Department of Physics, Tohoku University, Sendai 980-8578, Japan}
\author{Thom\'{a}s Fogarty}
\email{thomas.fogarty@oist.jp}
\author{Thomas Busch}
\email{thomas.busch@oist.jp}
\affiliation{Quantum Systems Unit, Okinawa Institute of Science and Technology Graduate University, Okinawa 904-0495, Japan}

\date{\today}

\begin{abstract}
We propose using the dynamical invariants, also known as the Lewis--Riesenfeld invariants, to speed-up the equilibration of a driven open quantum system. This allows us to reverse engineer the time-dependent master equation that describes the dynamics of the open quantum system and systematically derive a protocol that realizes a shortcut to equilibration. The method does not require additional constraints on the timescale of the dynamics beside the Born-Markov approximation and can be generically applied to boost single particle quantum engines significantly.
We demonstrate this with the damped harmonic oscillator, and show that our protocol can achieve high-fidelity control on shorter timescales than simple non-optimized protocols. We find that the system is heated during the dynamics to speed-up the equilibration, which can be considered as an analogue of the Mpemba effect in quantum control.
\end{abstract}

\maketitle

\section{Introduction}
Understanding and controlling open quantum systems is a major challenge for the exploration of quantum phenomena in the presence of dissipative effects, for the the deterministic preparation of quantum states, and for the development of quantum devices  \cite{Brif_2010,Vacanti_2014,Koch_2016,Kallush_2022,Koch_2022}. A particular interesting and relevant question is how to accelerate the equilibration of open quantum systems, which, for example, has applications in enhancing the performance of quantum heat engines \cite{Kosloff2014,Sai,Myers_2022,kurizki_kofman_2022}. Recently this question has attracted some attention and various techniques based on approaches such as optimal control \cite{Mukherjee_2013,Suri_2018,Vasco_2018,Xu_2022}, linear response theory \cite{Pancotti_2020}, techniques inspired by shortcuts to adiabaticity \cite{Villazon_2019,Alipour2020shortcutsto} and reverse engineering \cite{Dann_2019,Dupays_2020,Dann_2020} have been developed. Some of these  have also been successfully applied to design optimized quantum Carnot cycles \cite{Dann_2020b} or an optimized quantum Otto cycle \cite{Das_2020,Pedram_2023}, both of which show that a potential quantum advantage can be achieved.

The challenge of accelerating the equilibration of an open quantum system comes with the challenge of describing its dynamics in a practical way for control. Even in the well-established framework of using master equations to describe open quantum systems \cite{Breuer_2002}, their derivation can be a very difficult task for an underlying general time-dependent Hamiltonian. For example, a shortcut to equilibration (STE) in Refs.~\cite{Dann_2019,Dann_2020} was realized by deriving a non-adiabatic time-dependent master equation in the inertial limit  \cite{Dann_2018,Dann_2021}, which assumes small variations of the adiabatic parameter of the system. While this allows to obtain the Lindblad operators explicitly, the resulting driving protocol can be restricted.

In this work, we propose an STE realized by using the dynamical invariant (DI) to describe the dynamics of driven open quantum systems. Also known as the Lewis-Riesenfeld invariant, it was originally introduced by the latter to solve the time-dependent dynamics of closed quantum systems \cite{LR}. In the context of open quantum systems control, it has been recently shown that the DI can be used to quantify the decoherence and dissipation of a quantum system in the presence of an environment and therefore it can be used to design shortcuts to adiabaticity in order to drive an open quantum system toward a pure state \cite{cangemi2023control}. Moreover, the DI can also be used to derive a time-dependent master equation in a comprehensive way, without restrictions on the driving protocol, and which also gives a clear picture of the influence of the driving protocol on the dissipative effects \cite{Wu_2022,Wu_2015,Luo_2015}.

In this work we present a general formulation for a shortcut between the equilibrium states of the initial and final Hamiltonian of a driven open quantum system, by reverse engineering a time-dependent master equation that we derive based on the DI. To show the effectiveness of this approach, we apply this technique to the damped harmonic oscillator. This is a well-known model that can describe atomic and opto-mechanical systems, and is commonly used to explore the operation of quantum heat engines \cite{Rossnagel,Blickle_2012,Kosloff2017}. Moreover, dissipative harmonic oscillator systems have recently attracted additional attention as an interesting framework to describe Bose polaron systems \cite{Lampo2017bosepolaronas, Lampo_book}. These systems are promising platforms to explore quantum thermodynamics phenomena like heat transport between mesoscopic quantum gases \cite{Charalambous_2019} or thermometry \cite{Mehboudi_2019}. We show that our protocol outperforms non-optimized protocols, leading to much higher fidelity and shorter time to reach the target equilibrium state. Notably, the find that the effective temperature of the system in the middle of the dynamics needs to increase to achieve faster equilibration.

\section{Dynamical invariant based time-dependent master equation}\label{Sec2}
A driven, open quantum system is in general described by a Hamiltonian of the form
\begin{equation}\label{Hamiltonian}
    H(t)=H_{S}(t)+H_{B}+H_{I},
\end{equation}
where $H_{S}(t)$ is the time-dependent Hamiltonian of the system of interest and $H_{B}$ is the Hamiltonian of the environment. In our case, the environment corresponds to a thermal bath at a given temperature $T$ and thus its state is given by $\rho_{B}=Z_{B}^{-1}\exp\left(-\frac{H_{B}}{k_{B}T}\right)$. The final term describes the interaction between the system and the environment, $H_{I}=\sum_{k}A_{k}\otimes B_{k}$, where the $A_{k}$ act on the system and the $B_{k}$ act on the bath. To derive a solvable time-dependent master equation, we use the Born approximation
that assumes that the coupling strength is sufficiently weak to neglect correlations between the system and the bath during the dynamics. The state of the system and the bath can then be written as a product of the reduced states, $\rho(t)\approx\rho_{S}(t)\otimes\rho_{B}$. We also use the Markov approximation i.e.~we assume that the correlations inside the bath decay much faster than any other timescale of the system. This allows one to derive a master equation that is local in time. After tracing out the bath, we obtain a Redfield master equation in the interaction picture of the form \cite{Breuer_2002,REDFIELD19651}
\begin{equation}
\begin{split}
    \frac{d\Tilde{\rho}_{S}(t)}{dt}=-\frac{1}{\hbar^2}\sum_{k,l}\int_{0}^{\infty}B_{kl}(\tau)\left[\Tilde{A}_{k}(t),\Tilde{A}_{l}(t-\tau)\Tilde{\rho}_{S}(t)\right]\\
    -B_{lk}(-\tau)\left[\Tilde{A}_{k}(t),\Tilde{\rho}_{S}(t)\Tilde{A}_{l}(t-\tau)\right]d\tau,
\end{split}\label{redfield_me}
\end{equation}
where $B_{kl}(\tau)=\Tr_{B}\left(\Tilde{B}_{k}(\tau)B_{l}\rho_{B}\right)$ is the two-point correlation function of the bath and the tilde indicates operators in the interaction picture. In the presence of a time-dependent Hamiltonian, the evaluation of $\Tilde{A}_{k}(t)$ can be difficult. Indeed one needs to calculate the time evolution operator of the system $U_{S}(t)=T_{\leftarrow}\exp(-\frac{i}{\hbar}\int_{0}^{t}H_{S}(\tau)d\tau)$ (where $T_{\leftarrow}$ is the time-ordering operator) and this represents a technical challenge as the Hamiltonian at different times does usually not commute. An approximate solution for $U_{S}(t)$ can be found by using the inertial theorem that allows to reduce Eq.~\eqref{redfield_me} and derive a time-dependent master equation \cite{Dann_2018,Dann_2019,Dann_2021}. This solution is robust and valid when the inertial parameter is small which implies small variations of the adiabatic parameter of the system and therefore slow accelerations of the driving protocol.

However, the time-evolution operator can be found exactly by using the DI. Let us consider the closed dynamics of the system $i\hbar\partial_{t}\ket{\psi(t)}=H_{S}(t)\ket{\psi(t)}$. A DI of the Hamiltonian $H_{S}(t)$ is a Hermitian operator $I$ obeying \cite{LR}
\begin{equation}
    \frac{dI(t)}{dt}=\frac{\partial I(t)}{\partial t}+\frac{1}{i\hbar}[I(t),H_{S}(t)]=0.
\end{equation}
It has been shown that the solution of the closed dynamics can be written as a linear combination of the instantaneous eigenstates of the DI $\ket{\psi(t)}=\sum_{n}c_{n}e^{i\alpha_{n}(t)}\ket{\phi_{n}(t)}$ where the dynamical phases are given by $\alpha_{n}(t)=\frac{1}{\hbar}\int_{0}^{t}\bra{\phi_{n}(\tau)}i\hbar\frac{\partial}{\partial \tau}-H(\tau)\ket{\phi_{n}(\tau)}d\tau$. This property of the DI has previously been used to design shortcuts to adiabaticity in isolated systems \cite{STA,Xichen}. In that case a quantum system can be transported from a ground state to another one by ensuring that during the protocol, both the Hamiltonian and the DI commute initially and at the end of the protocol $\left[H_{S}(0),I(0)\right]=\left[H_{S}(t_{f}),I(t_{f})\right]=0$. The DI has also recently been used to design shortcuts to adiabaticity in open quantum systems \cite{cangemi2023control}. The eigenstates of the DI allow one to give an analytical expression of a map generated by a time-dependent Lindblad type master equation. This property is then exploited to construct a cost function that reduce the noise and dissipation due to the environment and therefore maximizes the fidelity with the target ground state during the protocol.

In contrast, here we derive an STE protocol and therefore the dynamics generated by such a protocol has to be highly non-unitary and needs to exploit the dissipative effects due to the coupling with  the environment. We use the DI to derive a time-dependent master equation describing the dynamics of a driven open quantum system, that we can then manipulate to find the desired protocol. Back to the Redfield equation \eqref{redfield_me}, we write the time-evolution operator of the system in terms of the eigenstates of the DI
\begin{equation}\label{te}
    U_{S}(t)=\sum_{n}e^{i\alpha_{n}(t)}\ket{\phi_{n}(t)}\bra{\phi_{n}(0)},
\end{equation}
which in turn allows us to calculate the operators acting on the system in the interaction picture as \cite{Wu_2022}
\begin{equation}
    \Tilde{A}_{k}(t)=\sum_{m,n}e^{i\left(\alpha_{n}(t)-\alpha_{m}(t)\right)}\bra{\phi_{m}(t)}A_{k}\ket{\phi_{n}(t)}F_{mn},
\end{equation}
with $F_{mn}=\ket{\phi_{m}(0)}\bra{\phi_{n}(0)}$. The operators can therefore be written as products of time-dependent scalar functions that contain the information on the driving protocol, and time-independent operators $F_{mn}$. Those operators are jump operators constructed with the DI eigenstate suggesting that the dissipative part of the open dynamics will involve transitions of the system between those states.

Focusing on the case of the time-dependent harmonic oscillator, the Hamiltonian of the system is given by $H_{S}(t)=\frac{p^{2}}{2m}+\frac{1}{2}m\omega(t)^{2}x^{2}$. A DI of this Hamiltonian is \cite{LR,Xichen}
\begin{equation}\label{invariant}
    I(t)=\frac{\left(b(t)p-m\Dot{b}(t)x\right)^{2}}{2m}+\frac{1}{2}m\omega_{0}^{2}\left(\frac{x}{b(t)}\right)^{2},
\end{equation}
where $\omega_{0}=\omega(0)$ and $b(t)$ corresponds to a  dimensionless scaling function that satisfies the Ermakov equation
\begin{equation}\label{ermakov}
    \ddot{b}(t)+\omega^{2}(t)b(t)=\frac{\omega_{0}^{2}}{b^{3}(t)}.
\end{equation}
We notice that the invariant has the structure of a harmonic oscillator with a constant frequency $\omega_{0}$, a position $x/b(t)$ and momentum $b(t)p-m\Dot{b}(t)x$. Thus the eigenstates can be obtained by using the standard ladder operators.

\section{Shortcut to equilibration for the damped harmonic oscillator}\label{Sec3}
\subsection{Time-dependent master equation}\label{Sec3a}
In what follows, we design an STE protocol for the time-dependent damped harmonic oscillator (DHO). The Hamiltonian of the bath is given by $H_{B}=\sum_{n}\hbar\omega_{n}\left(b_{n}^{\dagger}b_{n}+\frac{1}{2}\right)$ and the interaction is described in the rotating wave approximation by $H_{I}=\sum_{n}g_{n}\left(a^{\dagger}b_{n}+ab_{n}^{\dagger}\right)$, where the $g_{n}$ are constant coupling strengths between the particle and the $n$-th mode of the bath. By using the DI of Eq.~\eqref{invariant}, we describe the dynamics of the driven system by deriving the following Lindblad master equation in the interaction picture (see Appendix \ref{appA} for details) 
\begin{equation}\label{dho}
    \begin{split}
        &\frac{d\Tilde{\rho}_{S}(t)}{dt}=-\frac{i}{\hbar}\left[\Tilde{H}_{LS}(t),\Tilde{\rho}_{S}(t)\right] \\
        &+\frac{\abs{D(t)}^{2}}{2\hbar^{2}}\gamma_{+}(\Tilde{\omega}(t))\left(a_{I}\Tilde{\rho}_{S}(t)a_{I}^{\dagger}-\frac{1}{2}\{a_{I}^{\dagger}a_{I},\Tilde{\rho}_{S}(t)\}\right)\\
        &+\frac{\abs{D(t)}^{2}}{2\hbar^{2}}\gamma(\Tilde{\omega}(t))\left(a_{I}^{\dagger}\Tilde{\rho}_{S}(t)a_{I}-\frac{1}{2}\{a_{I}a_{I}^{\dagger},\Tilde{\rho}_{S}(t)\}\right),
    \end{split}
\end{equation}
where $\tilde{H}_{LS}(t)$ is the time-dependent Lamb shift in the interaction picture and $D(t)=b(t)+1/b(t)+i\dot{b}(t)/\omega_{0}$. The time-dependent decay rates characterizing the emission and absorption are given by $\gamma_{+}\left(\Tilde{\omega}(t)\right)=\pi J(\Tilde{\omega}(t))\left(1+n(\Tilde{\omega}(t))\right)$ and $\gamma\left(\Tilde{\omega}(t)\right)=\gamma_{+}(\tilde{\omega}(t))e^{-\frac{\hbar\tilde{\omega}(t)}{k_{B}T}}$, with $J(\omega)=\sum_{n}g_{n}^{2}\delta(\omega-\omega_{n})$ being the bath spectral density function and $n(\omega)=(e^{\frac{\hbar\omega}{k_{B}T}}-1)^{-1}$ the Planck distribution. The time-dependent decay rates $\gamma$ and $\gamma_{+}$ depend on an effective time-dependent Bohr frequency given by $\tilde{\omega}(t)=\omega_{0}/b(t)^{2}$. This means that the scaling function $b(t)$ sets the frequency at which the absorption and emission of excitations from the particle occur during the dynamics and therefore how the decay rates evolve. The scaling function is also related to the trap frequency $\omega(t)$ through the Ermakov equation \eqref{ermakov}, which means that the trap frequency and the decay rates are related to each other through the scaling function. This implies, as we will show later, that we only need to find the scaling function and control the trap frequency in order to realize the STE protocol successfully, and no particular manipulation of the decay rates is needed, as the driving protocol indirectly controls the decay rates.

The Lindblad operators $a_{I}$ and $a_{I}^{\dagger}$ correspond respectively to the annihilation and creation operator of the invariant at $t=0$. However, if one considers a driving protocol with a continuous start from the initial Hamiltonian $[H_{S}(0),I(0)]=0$, we recover the creation and annihilation operators of the particle, $a_{I}=a$ ($a_{I}^{\dagger}=a^{\dagger}$). The range of validity of the master equation imposes constrains on the driving protocol $\omega(t)$. Indeed the Markov approximation implies that the typical timescale of the system $\tau_{S}$ is much larger than the timescale at which the bath two-point correlation functions decay $\tau_{B}$ . The typical timescale of the harmonic oscillator is $\tau_{S}\sim\omega(t)^{-1}$ and the decay time of the bath two-point correlation functions is related to the cut-off of the bath's spectral density function $\Lambda$ as $\tau_{B}\sim\Lambda^{-1}$. Therefore the Markov approximation is valid if $\omega(t)^{-1}\gg\Lambda^{-1}$. We have also used the Born approximation that allows one to make a second order approximation in the coupling strength $g$ between the system and the bath and thus to start from the Redfield equation \eqref{redfield_me}. One can show that the next leading order correction term in the perturbation is negligible if $g\tau_{B}\ll1$ i.e $g\Lambda^{-1}\ll1$ \cite{Albash_2012}.

Additionally we used a secular approximation that implies $\int_{0}^{t}\tilde{\omega}(\tau)d\tau\gg\tilde{\omega}(t)\tau_{B}$ in order to derive a master equation in Lindblad form and to ensure that the state of the system remains positive during the dynamics. Finally a last approximation has been made that can be formulated as $\tau_{B}\ll\tau_{D}$ where we introduced a driving timescale $\tau_{D}=\min_{t}\abs{D(t)/\dot{D}(t)}$ \cite{Dann_2018}. This approximation is not necessary in order to derive the master equation, however it allows to simplify its form and thus the reverse-engineering for the STE. This approximation implies that the time window in which the change of $D$ and therefore the change of the scaling function $b$ becomes significant for the system has to be large enough compared to the characteristic time of the bath $\tau_{B}$. This approximation is consistent with the Markov approximation (see Appendix \ref{appA} for more details).

\subsection{Formulation of the shortcut protocol and reverse engineering}\label{Sec3b}
We design a shortcut protocol for an isothermal stroke where the trap frequency of the system is driven from $\omega_0$ to $\omega_f$ while coupled to a bath at temperature $T$. In principle the system needs to be driven slowly in order to remain at equilibrium with the bath at all times during the process. In order to speed up this process our STE protocol will consist of mimicking the stroke at finite time by connecting the initial and final desired equilibrium states. Initially the particle is at equilibrium with the frequency $\omega(0)=\omega_{0}$ i.e $\rho_{S}(0)=Z_{0}^{-1}e^{-H_{S}(0)/k_{B}T}$. The final desired state is an equilibrium state of the particle with the bath at the trapping frequency $\omega(t_{f})=\omega_{f}$ i.e $\rho_{S}(t_{f})=Z_{f}^{-1}e^{-H_{S}(t_{f})/k_{B}T}$, where $t_{f}$ is the duration of the protocol. We focus on the compression stroke $\omega_{f}>\omega_{0}$ but the expansion can also be done in the same way. 

We know that the state of the particle will be Gaussian during the dynamics since the Hamiltonian is quadratic \cite{Eisert_2003}. Therefore it can be fully determined by the expectation values $\langle a^{\dagger}a\rangle(t)=\Tr(a^{\dagger}a\rho_{S}(t))$ and $\langle a^{2}\rangle(t)=\Tr(a^{2}\rho_{S}(t))$ characterizing the excitation and the squeezing of the particle. The equations describing their evolution during the driving protocol are obtained in the interaction picture from the master equation \eqref{dho} as
\begin{align}
    &\frac{d\langle\Tilde{a}^{\dagger}\Tilde{a}\rangle}{dt}=\frac{\pi}{2\hbar^{2}}\abs{D(t)}^{2}J(\tilde{\omega}(t))\left(n(\Tilde{\omega}(t))-\langle\Tilde{a}^{\dagger}\Tilde{a}\rangle\right),\label{nt}\\
    &\frac{d\langle\tilde{a}^{2}\rangle}{dt}=-\frac{\pi}{2\hbar^{2}}\abs{D(t)}^{2}J(\Tilde{\omega}(t))\langle\Tilde{a}^{2}\rangle.\label{a2}
\end{align}
Since the particle is initially in a Gibbs state with $\langle \tilde{a}^{2}\rangle(0)=0$, it follows from Eq.~\eqref{a2} that $\langle \tilde{a}^{2}\rangle=0$ at any time during  the dynamics. Thus, the protocol is  described by the differential equation \eqref{nt} alone. Furthermore, the initial and target state are equilibrium states which implies $\langle\Tilde{a}^{\dagger}\Tilde{a}\rangle(0)=(e^{\frac{\hbar\omega_{0}}{k_{B}T}}-1)^{-1}$, $\langle\Tilde{a}^{\dagger}\Tilde{a}\rangle(t_{f})=(e^{\frac{\hbar\omega_{f}}{k_{B}T}}-1)^{-1}$ and $\frac{d\langle\Tilde{a}^{\dagger}\Tilde{a}\rangle(0)}{dt}=\frac{d\langle\Tilde{a}^{\dagger}\Tilde{a}\rangle(t_{f})}{dt}=0$. In addition to these boundary conditions, we impose $\frac{d^{2}\langle\Tilde{a}^{\dagger}\Tilde{a}\rangle(0)}{dt^{2}}=\frac{d^{2}\langle\Tilde{a}^{\dagger}\Tilde{a}\rangle(t_{f})}{dt^{2}}=0$ to ensure a smooth evolution of the system between the initial state and the target state.

\begin{figure}
\includegraphics[width=\linewidth]{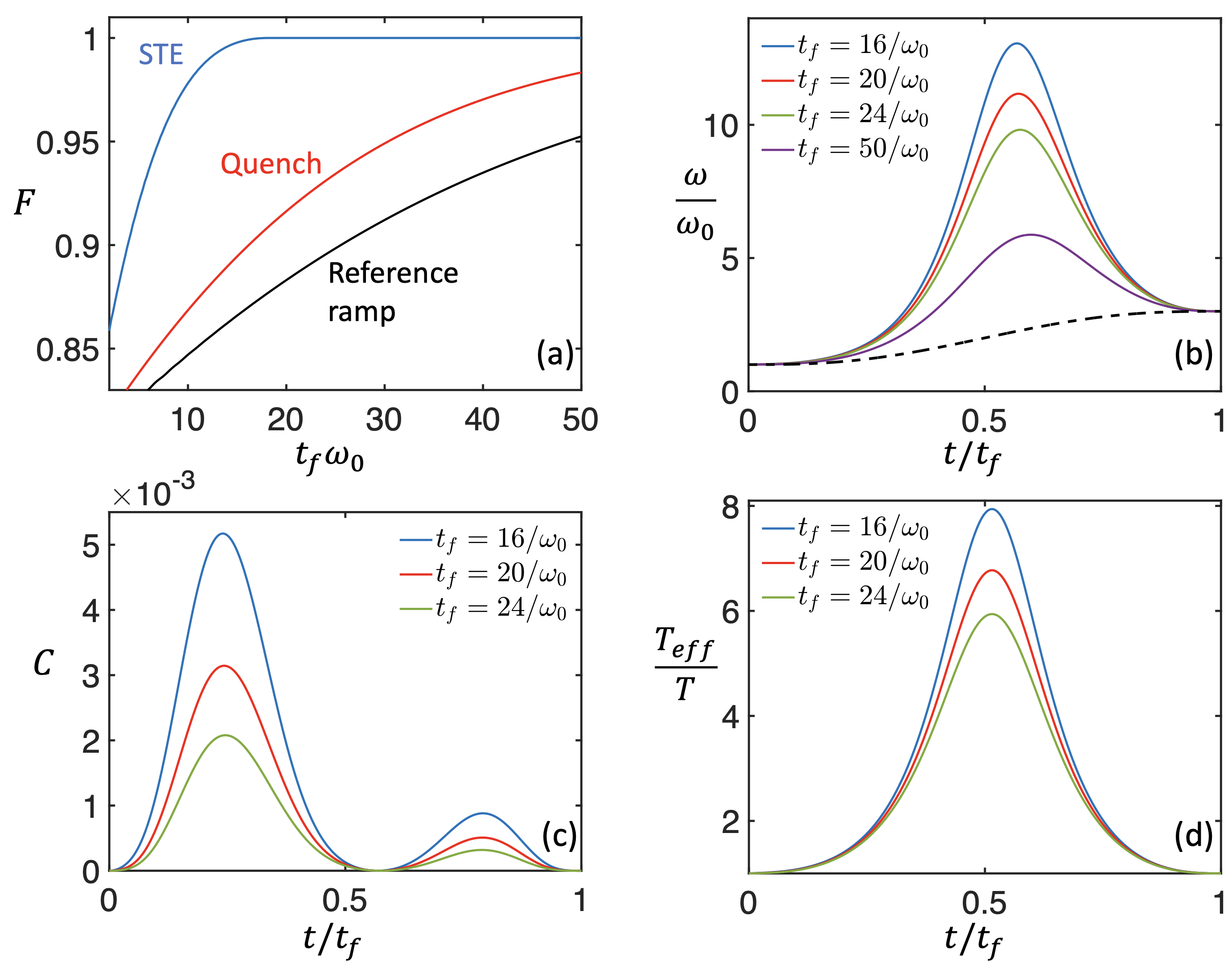}
\caption{(a) Fidelity between the final state and the target state as a function of $t_{f}$. (b) Profile of the trap frequency for the STE protocol as a function of time for different protocol durations. The black dashed line shows the reference ramp. (c) Coherence generated during the dynamics of the STE as a function of time. (d) Effective temperature of the particle during the STE protocol as a function of time. The final compression is $\omega_{f}=3\omega_{0}$ and the bath temperature is $T=\hbar\omega_{0}/k_{B}$. The spectral density function is an Ohmic distribution with an abrupt cut-off $J(\omega)=\gamma\omega\Theta(\Lambda-\omega)$ with $\gamma=\hbar^{2}/500$ and $\Lambda=100\omega_{0}$, and the number of particles in the bath is $N=600$.}\label{plot_dho}
\end{figure}

The protocol can now be found by reverse-engineering Eq.~\eqref{nt} to obtain the scaling function $b(t)$, which in turn allows one to obtain the trap frequency from the Ermakov equation \eqref{ermakov} \cite{Xichen}. When it is doable, reverse-engineering can be an interesting alternative to find a control protocol compared to optimal control methods, from a computational point of view. Indeed while optimal control methods are robust in finding a control protocol for driving a system to a specific state \cite{Koch_2022}, they rely on minimization schemes where the dynamics needs to be solved at each iteration in order to compute a given cost function, which is more demanding.

We assume that the evolution of $\langle a^{\dagger}a\rangle$ is given by a 5-th order polynomial function that ensures that the boundary conditions obtained above are satisfied. The boundary conditions on $\langle\Tilde{a}^{\dagger}\Tilde{a}\rangle$ combined with Eq.~\eqref{nt} also imply $b(0)=1$, $b(t_{f})=\sqrt{\omega_{0}/\omega_{f}}$ and $\dot{b}(0)=\dot{b}(t_{f})=0$. We also obtain additional boundary conditions from the Ermakov equation $\ddot{b}(0)=\ddot{b}(t_{f})=0$. We can thus consider a 6-th order polynomial ansatz for the scaling function $b(t)=\sum_{n=0}^{6}a_{n}(t/t_{f})^{n}$, in which the first 6 coefficients allow to satisfy the above boundary conditions. The 6-th order term can then ensure that the scaling function connects the initial state and the target state through Eq.~\eqref{nt}. The coefficient $a_{6}$ is simply found by maximizing the fidelity between the target state and the state of the particle at the end of the protocol. Using this ansatz for the scaling function allows to avoid the divergence issues that one can encounter when solving Eq.~(9), where at $t=0$ and $t=t_{f}$ the average excitation $\langle a^{\dagger}a\rangle$ and the Planck distribution $n(\tilde{\omega}(t))$ converge to the same values.

\subsection{Properties of the shortcut}\label{Sec3c}
To quantify the performance of the shortcut, we calculate the fidelity between the target state $\rho_T$ and the state of the particle at the end of the protocol
\begin{equation}
    F(\rho_{S}(t_{f}),\rho_{T})=\Tr\left(\sqrt{\sqrt{\rho_{T}}\rho_{S}(t_{f})\sqrt{\rho_{T}}}\right)^{2}.
\end{equation}
The STE protocol has been found using the time-dependent master equation \eqref{dho} and, as long as the dynamics is within its scope of description, the fidelity will be always one. However the approximations that we considered in order to derive the master equation might be violated depending on the time $t_{f}$, the driving protocol $\omega(t)$ that we found and the properties of the bath (number of particles, spectral density function $J$). For these reasons, we have tested the STE protocol by solving  the dynamics of the DHO model exactly. Since the states of the particle and the bath remain Gaussian, we can use an efficient numerical method to solve the dynamics of the total Hamiltonian \eqref{Hamiltonian} during the protocol. The method consists of time-evolving the covariance matrix of the system plus the bath, which corresponds to a $(2N+2)\times (2N+2)$ matrix (where $N$ is the number of particles in the bath), with the Heisenberg equations of motion \cite{Rivas_2010}. This allows us to see when the validity of the master equation actually breaks down for a given set of bath parameters and therefore when the shortcut does not work anymore. 

To demonstrate the benefit of our shortcut, we compare it with two simple protocols: the sudden quench $\omega_{q}(t>0)=\omega_{f}$ and a reference ramp described by a polynomial function  $\omega_{r}(t)=\omega_{0}+10\Delta\omega(t/t_{f})^{3}-15\Delta\omega(t/t_{f})^{4}+6\Delta\omega(t/t_{f})^{5}$ with $\Delta\omega=\omega_{f}-\omega_{0}$. For the DHO, the quench protocol is characterized by an asymptotic exponential convergence of the fidelity to one as $F\approx1-e^{-t_{f}/\tau_{q}}$ where $\tau_{q}$ is a characteristic time related to the decay rates \cite{Dann_2019}. 

The fidelity as a function of $t_{f}$ is shown in Fig.~\ref{plot_dho}(a). The STE outperforms the sudden quench and the reference ramp that also shows an asymptotic exponential behavior but with a longer characteristic time than the sudden quench. However, one can see that for short durations, the fidelity for the STE protocol collapses because at shorter times the description provided by the master equation deviates from the exact dynamics. The fidelity reaches approximately the value 0.999 around $t_{f}\approx16/\omega_{0}$ and then keeps increasing to one.

In order to obtain physical insights on the STE protocol the profile of the trap frequency $\omega(t)$ is shown in Fig.~\ref{plot_dho}(b) for different protocol durations $t_{f}$ and compared to the reference ramp (black dashed line). We see that the profiles are quite different: while for the reference ramp the frequency increases monotonically toward $\omega_{f}$, in the STE protocol the particle is driven to large trap frequency values at intermediate times before decreasing to reach the final frequency at $t=t_f$. Notably, for faster protocols the particle is driven to higher frequency values. Inversely, for larger $t_f$, the amplitude of the trap frequency decreases and we observe that the STE protocol gets closer to the reference ramp when $t_{f}$ approaches the adiabatic limit.

Such a trap frequency profile necessarily implies non-equilibrium features in the dynamics. To quantify them, we calculate the coherence in the system during the STE, which has been suggested to play a key role in the control of open quantum systems \cite{Dann_2020b}. We define it as the change of entropy between the diagonal part of the state and the full density matrix in the instantaneous eigenenergy basis \cite{Baumgratz_2014}
\begin{equation}
    C(t)=S\left(\rho_{diag}(t)\right)-S(\rho_{S}(t)),
\end{equation}
where $S(\rho)=-\Tr(\rho\log(\rho))$ is the von Neumann entropy. The coherence is shown in Fig.~\ref{plot_dho}(c). One can see that its profile reflects the results we have shown before: coherence is generated in the system when the trap frequency is changing. It vanishes when the maximum trap frequency is reached, and increases again when the trap frequency decreases toward the final value. There is no coherence remaining in the system at the end of the protocol since the particle reaches the target Gibbs state. Moreover the amount generated during the dynamics increases for faster protocols. This also allows to explain why the quench protocol works better than the reference ramp. Indeed, coherence causes transitions of the particle between its eigenstates and can be used as a catalysis that helps to accelerate the thermalization of an open quantum system. A controlled manipulation of coherence allows to reach the new equilibrium state and this is what the STE achieves. Even though the non-optimal protocols do not reach the target state, the sudden quench will always perform better than a smooth ramp that generates much less non-adiabatic excitations. Notably, this is in contrast with shortcuts to adiabaticity protocols which require the suppression of such excitations to reach the desired target state \cite{STA}. Even though we use a different description of the dynamics of the system, the generation of coherence we observe is in agreement with the STE protocol obtained with the non-adiabatic master equation derived in Ref.~\cite{Dann_2018}. This confirms its crucial role in achieving fast thermalization processes in driven open quantum systems \cite{Dann_2020b}.

After using the DI to derive the master equation, we now show and exploit some of its remarkable properties in order to characterize the state of the particle during the dynamics of the STE further. Indeed based on the knowledge of the dynamics established in the formulation of the shortcut protocol, we can show that during the dynamics, the state of the particle can be written in the interaction picture as $\Tilde{\rho}_{S}(t)=Z(t)^{-1}\sum_{n}e^{-\epsilon(t)n}\ket{\phi_{n}(0)}\bra{\phi_{n}(0)}$ with 
\begin{equation}\label{Planck_eff}
    \langle\Tilde{a}^{\dagger}\Tilde{a}\rangle(t)=(e^{\epsilon(t)}-1)^{-1},
\end{equation}
and the effective partition function is given by $Z(t)=(1-e^{-\epsilon(t)})^{-1}$. Back to the Schr\"odinger picture the Hamiltonian is always diagonal in the instantaneous eigenbasis of the DI $\rho_{S}(t)=Z(t)^{-1}\sum_{n}e^{-\epsilon(t)n}\ket{\phi_{n}(t)}\bra{\phi_{n}(t)}$. Therefore we have shown that similarly to the case of closed systems, the eigenstates of the DI give us the states that the system will explore for a given protocol. This allows us to simply define an effective temperature of the system during the dynamics as 
\begin{equation}
    T_\text{eff}(t)=\frac{\hbar\omega(t)}{k_{B}\epsilon(t)}.
\end{equation}
Defining a temperature for non-equilibrium sates can be useful in many cases \cite{JCasas-Vázquez_2003,PUGLISI20171} and is defined in our case from the effective Planck distribution Eq.~\eqref{Planck_eff}. Therefore besides characterizing how far from equilibrium the system is during the protocol, it is also related to both the energy and entropy of the particle. A high effective temperature is therefore associated with a state of high energy and mixedness. The effective temperature of the particle during the STE is shown in Fig.~\ref{plot_dho}(d). It deviates significantly from the bath temperature before returning to it at the end of the protocol. More interestingly, the particle is driven to states that are effectively hotter, and the faster the shortcut is, the hotter the state of the particle is. While the shortcut is designed for an isothermal compression, which corresponds to a cooling process, the strategy adopted by the STE actually consists of warming up the particle in order to cool it down faster. This is  reminiscent of the Mpemba effect \cite{Mpemba_1969}, an empirical phenomenon where a hot liquid can freeze faster than a cold liquid. Recently, the Mpemba effect has been discussed and predicted for a quantum dot coupled to two reservoirs \cite{Chatterjee_2023}. Here, we observe a similar feature to the thermal Mpemba effect in the context of driven open quantum systems, with the same behavior being previously observed in the effective temperature for an STE of a two-level system \cite{Dann_2020}.

\section{Conclusion}\label{Concl}
In the present work, we have presented results that contribute to the improvement of controlling driven open quantum systems and the  performance of quantum heat engines. We have shown that dynamical invariants can be a powerful tool for describing and accelerating the equilibration for these systems, and that besides the Born-Markov approximation, it lacks additional restrictions on the timescale of the dynamics. Our work also brings new physical interpretations of the dynamical invariant. Indeed the scaling function $b(t)$, that fully characterizes the invariant, sets both the driving protocol and the decay rates. This allows to derive protocols that modify both the unitary part and the dissipative part of the dynamics at the same time by only changing the trapping potential. Therefore an experimental implementation is feasible as one only needs to control the trapping potential, which is in contrast with  the method proposed in \cite{Alipour2020shortcutsto} where one needs to find and implement both the trapping potential and the decay rates in a specific way. The STE protocol is characterized by a manipulation of the coherence that drives the particle to hotter states when designed for an isothermal compression. Our observation resonates with the thermal Mpemba effect and a rigorous formulation of this phenomenon in the context of driven open quantum systems would be an interesting direction to take in the future.  

The STE protocol comes with dissipated work that is generated due to irreversibility \cite{Landi_2021} and that can actually limit the boost on the performance of quantum heat engines for small timescales \cite{Dann_2020b}. Therefore it is an interesting avenue to design shortcuts by minimizing the irreversibility and address the question whether the geometric bound derived for the dissipated work \cite{Salamon_1983,Scandi2019thermodynamiclength,Li_2022} can be reached with our shortcut. While we have considered the isothermal stroke in this work, one can consider different strokes, or start from non-equilibrium states. It would be also possible to design shortcuts for applications other than ones thermodynamics. For example, 
one could design fast and robust protocols for quantum gates \cite{Kallush_2022,turyansky2023inertial}. Also recently, a similar approach has been used to quickly generate entangled states in a double two-level system \cite{Ma_2023}. 

Another important extension would be to go beyond the single particle problem and optimize the equilibration of interacting many-body states. The task is challenging because obtaining a DI for complex quantum systems is generally difficult. However, one could consider two particles with short-range interactions where the problem can be split into two non-interacting particles \cite{Busch98}, or a bosonic gas in the hardcore Tonks--Girardeau limit where the system can be described by spin polarized fermions through the Bose-Fermi mapping theorem \cite{Girardeau}. Beside having well-known analytical results, both systems have also shown enhanced performances compared to non-interacting quantum engines \cite{Jaramillo_2016,Fogarty_2020,me_2023}, paving the way for fully-optimized many-body quantum heat engines.

Furthermore, an exact DI has recently been found for interacting spin systems \cite{LI2023229}, while a general method to construct DIs for complex systems has been proposed in Ref.~\cite{Ponte_2018}. Extending our STE approach to more systems is therefore within reach. Finally another promising direction to take would be to first use our approach to find an STE protocol for a non-interacting model and then add a correction term that improves the fidelity between the target state and the final state when adding interactions. Two ways can be considered to find the correction term. The first one is analytical and consists of doing a quadratic expansion of the
fidelity similar to the ``enhanced shortcut to adiabaticity” method \cite{Whitty_2020}. The second way is numerical and consists of getting the correction term by using optimal control methods, similarly to the
``counterdiabatic optimized local driving” method that has been recently proposed to improve STAs in closed quantum systems \cite{Cepaite_2023}.

\acknowledgements
This work was supported by the Okinawa Institute of Science and Technology Graduate University, and used the computing resources of the Scientific Computing and Data Analysis section at OIST. The authors also acknowledge support from a SHINKA grant from OIST and Tohoku University. SE is supported by JSPS KAKENHI Grant Numbers JP21H00116 and JP22K03492. TF acknowledges support from JSPS KAKENHI Grant Number JP23K03290. TF and TB are also supported by JST Grant Number JPMJPF2221.

\bibliography{main}

\appendix
\section{Time-dependent master equation of the damped harmonic oscillator}\label{appA}
We start from the Redfield equation given by Eq.~\eqref{redfield_me} applied to the damped harmonic oscillator
\begin{equation}\label{dho_red}
\begin{split}
    &\frac{d\Tilde{\rho}_{S}(t)}{dt}=-\frac{1}{\hbar^2}\int_{0}^{\infty}B_{12}(\tau)\left[\tilde{a}^{\dagger}(t),\tilde{a}(t-\tau)\tilde{\rho}_{S}(t)\right]\\
    &-B_{21}(-\tau)\left[\tilde{a}^{\dagger}(t),\tilde{\rho}_{S}(t)\tilde{a}(t-\tau)\right]\\
    &+B_{21}(\tau)\left[\tilde{a}(t),\tilde{a}^{\dagger}(t-\tau)\tilde{\rho}_{S}(t)\right]\\
    &-B_{12}(-\tau)\left[\tilde{a}(t),\tilde{\rho}_{S}(t)\tilde{a}^{\dagger}(t-\tau)\right]d\tau,
\end{split}
\end{equation}
where the bath two-point correlation functions are given by
\begin{equation}
    \begin{split}
        B_{12}(\tau)&=\sum_{n}g_{n}^{2}\Tr_{B}(\tilde{b}_{n}(\tau)b_{n}^{\dagger}\rho_{B})=\sum_{n}e^{-i\omega_{n}\tau}g_{n}^{2}(1+n(\omega_{n}))\\
        &=\int_{0}^{\infty}e^{-i\omega\tau}J(\omega)(1+n(\omega))d\omega,\\
        B_{21}(\tau)&=\sum_{n}g_{n}^{2}\Tr_{B}(\tilde{b}_{n}^{\dagger}(\tau)b_{n}\rho_{B})=\sum_{n}e^{i\omega_{n}\tau}g_{n}^{2}n(\omega_{n})\\
        &=\int_{0}^{\infty}e^{i\omega\tau}J(\omega)n(\omega)d\omega.
    \end{split}
\end{equation}

Now we need to evaluate the ladder operators of the particle in the interaction picture. Since they are conjugate, we can just focus on the annihilation operator, which we can write with the position and momentum operators in the interaction picture as
\begin{equation}
    \tilde{a}(t)=\sqrt{\frac{m\omega_{0}}{2\hbar}}\left(\tilde{x}(t)+i\frac{\tilde{p}(t)}{m\omega_{0}}\right).
\end{equation}

We use the invariant \eqref{invariant} to evaluate the operator in the interaction picture. Since the invariant is a harmonic oscillator with a position $x/b(t)$ and momentum $\Pi=b(t)p-m\dot{b}(t)x$, we can express the position and momentum of the particle in terms of the instantaneous ladder operators of the invariant that we denote $a_{I_{t}}$ and $a_{I_{t}}^{\dagger}$
\begin{equation}\label{A4}
    \begin{split}
        &x=b(t)\sqrt{\frac{\hbar}{2m\omega_{0}}}(a_{I_{t}}+a^{\dagger}_{I_{t}}),\\
        &p=\frac{\Pi}{b(t)}+m\dot{b}(t)\frac{x}{b(t)}=\sqrt{\frac{\hbar m\omega_{0}}{2}}\left(C(t)a_{I_{t}}+C^{*}(t)a_{I_{t}}^{\dagger}\right),
    \end{split}
\end{equation}
with the complex function 
\begin{equation}
   C(t)=\frac{\dot{b}(t)}{\omega_{0}}-\frac{i}{b(t)}. 
\end{equation}

We deduce that the annihilation operator of the harmonic oscillator in the Schr\"odinger picture is related to the ladder operators of the invariant through the following Bogoliubov transformation
\begin{equation}
    a=\frac{1}{2}\left(D_{1}(t)a_{I_{t}}+D_{2}^{*}(t)a^{\dagger}_{I_{t}}\right),
\end{equation}
where
\begin{equation}
        D_{1,2}(t)=b(t)\pm\frac{1}{b(t)}\pm i\frac{\dot{b}(t)}{\omega_{0}}.
\end{equation}

We now need to calculate the ladder operators of the invariant in the interaction picture. This is easily done by using the time-evolution operator written with the eigenstates of the invariant (Eq.~\eqref{te}) and using the expression of the dynamical phase for the harmonic oscillator $\alpha_{n}(t)=-\omega_{0}(n+1/2)\int_{0}^{t}1/b(\tau)^{2}d\tau$
\begin{equation}
\begin{split}
    \tilde{a}_{I_{t}}(t)&=\sum_{n,m}e^{i(\alpha_{m}(t)-\alpha_{n}(t))}\ket{\phi_{n}(0)}\bra{\phi_{n}(t)}a_{I_{t}}\ket{\phi_{m}(t)}\bra{\phi_{m}(0)}\\
    &=\sum_{n,m}e^{i(\alpha_{m}(t)-\alpha_{n}(t))}\sqrt{m}\delta_{n,m-1}\ket{\phi_{n}(0)}\bra{\phi_{m}(0)}\\
    &=e^{-i\varphi(t)}\sum_{n}\sqrt{n+1}\ket{\phi_{n}(0)}\bra{\phi_{n+1}(0)}=e^{-i\varphi(t)}a_{I_{0}},
\end{split}
\end{equation}
where the phase $\varphi$ is given by

\begin{equation}
    \varphi(t)=\int_{0}^{t}\frac{\omega_{0}}{b(\tau)^{2}}d\tau=\int_{0}^{t}\tilde{\omega}(\tau)d\tau.
\end{equation}

For the next we will use $a_{I}$ ($a^{\dagger}_{I}$) instead of $a_{I_{0}}$ ($a^{\dagger}_{I_{0}}$) to denote the ladder operators of the invariant at $t=0$. Now we can obtain an explicit expression of the annihilation operator of the particle in the interaction picture
\begin{equation}\label{a_int}
    \tilde{a}(t)=\frac{1}{2}\left(D_{1}(t)e^{-i\varphi(t)}a_{I}+D_{2}^{*}(t)e^{i\varphi(t)}a_{I}^{\dagger}\right).
\end{equation}

We can insert Eq.~\eqref{a_int} in the Redfield equation \eqref{dho_red}. We only explicitly write it down for the first commutator in the right-hand side of Eq.\eqref{dho_red} since the same treatment can be straightforwardly done for the other terms. After expanding the commutator, we obtain

\begin{widetext}
\begin{equation}\label{a10}
\begin{split}
     &[\tilde{a}^{\dagger}(t),\tilde{a}(t-\tau)\tilde{\rho}_{S}(t)]=\frac{1}{4}\left(e^{-i(\varphi(t)-\varphi(t-\tau))}D_{2}(t)D_{2}^{*}(t-\tau)[a_{I},a_{I}^{\dagger}\tilde{\rho}_{S}(t)]+e^{-i(\varphi(t)+\varphi(t-\tau))}D_{2}(t)D_{1}(t-\tau)[a_{I},a_{I}\tilde{\rho}_{S}(t)]\right.\\
     &\left.+e^{i(\varphi(t)+\varphi(t-\tau))}D_{1}^{*}(t)D_{2}^{*}(t-\tau)[a_{I}^{\dagger},a_{I}^{\dagger}\tilde{\rho}_{S}(t)]+e^{i(\varphi(t)-\varphi(t-\tau))}D_{1}^{*}(t)D_{1}(t-\tau)[a_{I}^{\dagger},a_{I}\tilde{\rho}_{S}(t)]\right).
\end{split}
\end{equation}
\end{widetext}

The integral in the Redfield equation is dominated by the bath two-point correlation function that rapidly decays with a characteristic time $\tau_{B}$. The decay time is given by the cut-off of the bath $\tau_{B}\sim\Lambda^{-1}$. Based on the Markov approximation, the decay time must be much smaller than the typical timescale of the system given by $\omega(t)^{-1}$ i.e $\tau_{B}\ll\omega(t)^{-1}$. We can thus use the first order approximation of the phase in the integral $\varphi(t-\tau)\approx\varphi(t)-\tilde{\omega}(t)\tau$. We also make the zero-th order approximation $D_{i}(t-\tau)\approx D_{i}(t)$ meaning that the variations of the scaling function $b(t)$ and its derivative are negligible in the time window $[0,\tau_{B}]$. Formally, it implies $\tau_{B}\ll\abs{\frac{D_{i}(t)}{\dot{D}_{i}(t)}}$. This approximation can be reformulated as $\tau_{B}\ll\tau_{D}$ where we introduce a driving timescale $\tau_{D}=\min_{i,t}\abs{\frac{D_{i}(t)}{\dot{D}_{i}(t)}}$ \cite{Dann_2018}. Let us remark that this approximation is not necessary to derive the master equation, however it allows to simplify the reverse-engineering for the shortcut.

The last approximation we will use is the secular approximation. We neglect the non-secular terms, to derive a master equation in Lindblad form and ensure that the state of the system remains physical. This means that the non-secular contributions contain fast oscillating terms that average to zero. This implies $\varphi(t)+\varphi(t-\tau)\gg\varphi(t)-\varphi(t-\tau)$. By using a first order expansion and the Markov approximation, we obtain $\varphi(t)\gg\tilde{\omega}(t)\tau_{B}$ i.e $\int_{0}^{t}\tilde{\omega}(\tau)d\tau\gg\tilde{\omega}(t)\tau_{B}$. 

Taking account of the different approximations in Eq.~\eqref{a10}, we obtain

\begin{widetext}
\begin{equation}
\int_{0}^{\infty}B_{12}(\tau)\left[\tilde{a}^{\dagger}(t),\tilde{a}(t-\tau)\tilde{\rho}_{S}(t)\right]d\tau\approx\frac{\abs{D_{2}(t)}^{2}}{4}\int_{0}^{\infty}B_{12}(\tau)e^{-i\tilde{\omega}(t)\tau}d\tau[a_{I},a_{I}^{\dagger}\tilde{\rho}_{S}(t)]+\frac{\abs{D_{1}(t)}^{2}}{4}\int_{0}^{\infty}B_{12}(\tau)e^{i\tilde{\omega}(t)\tau}d\tau[a_{I}^{\dagger},a_{I}\tilde{\rho}_{S}(t)].
\end{equation}
\end{widetext}

The integrals can be calculated by using the well-known result $\int_{0}^{\infty}e^{i\omega\tau}d\tau=\pi\delta(\omega)+iP(1/\omega)$ where $P$ denotes the principal value. After combining the different terms and few lines of algebra, we obtain the time-dependent master equation for the damped harmonic oscillator given by Eq.~\eqref{dho}, where we replaced $D_{1}(t)$ by $D(t)$. The time-dependent Lamb shift is given by
\begin{equation}
\begin{split}
    &\tilde{H}_{LS}(t)=\frac{\hbar}{4}\left(\abs{D_{1}(t)}^{2}P\int_{0}^{\infty}\frac{J(\omega)}{\tilde{\omega}(t)-\omega}d\omega\right.\\
    &\left.-\abs{D_{2}(t)}^{2}P\int_{0}^{\infty}\frac{J(\omega)}{\tilde{\omega}(t)+\omega}d\omega \right).
\end{split}
\end{equation}

\end{document}